\documentclass{article}

\usepackage{microtype}
\usepackage{graphicx}
\usepackage{subcaption}
\usepackage{booktabs} % for professional tables

\usepackage{hyperref}

\usepackage[preprint]{icml2026}

\usepackage{amsmath}
\usepackage{amssymb}
\usepackage{mathtools}
\usepackage{amsthm}

\usepackage[capitalize,noabbrev]{cleveref}

\theoremstyle{plain}

\theoremstyle{definition}

\theoremstyle{remark}

\usepackage{graphicx}
\usepackage{amsmath}
\usepackage{amssymb}
\usepackage{array}
\usepackage{bm}
\usepackage{booktabs}
\usepackage{cancel}
\usepackage{color}
\usepackage{euscript, latexsym}
\usepackage{enumerate}
\usepackage[shortlabels]{enumitem}
\usepackage{stmaryrd}
\usepackage{subcaption}
\usepackage{xcolor}
\usepackage{xurl}
\usepackage{microtype}

\newcommand{\abs}[1]{\left\vert #1 \right\vert}
\newcommand{\norm}[1]{\left\vert\left\vert #1 \right\vert\right\vert}
\newcommand{\paren}[1]{\left( #1 \right)}
\newcommand{\braces}[1]{\left\{ #1 \right\}}
\newcommand{\bracket}[1]{\left[ #1 \right]}
\newcommand{\ket}[1]{\left\vert #1 \right\rangle}

\newcommand{\ip}[1]{\left\langle #1 \right\rangle}

\newcommand{\E}{\mathbb{E}}
\newcommand\numberthis{\addtocounter{equation}{1}\tag{\theequation}}
\usepackage[textsize=tiny]{todonotes}

\icmltitlerunning{Neural Fourier Surrogates for Data Reuploading Quantum Neural Networks}

\begin{document}

\twocolumn[
  \icmltitle{Neural Fourier Surrogates for Data Reuploading Quantum Neural Networks}



  \icmlsetsymbol{equal}{*}

  \begin{icmlauthorlist}
    \icmlauthor{Oliver Knitter}{yyy}
    \icmlauthor{Jonathan Mei}{yyy}
    \icmlauthor{Sang Hyub Kim}{yyy}
    \icmlauthor{Chi Chen}{yyy}
    \icmlauthor{Masako Yamada}{yyy}
    \icmlauthor{Martin Roetteler}{yyy}
  \end{icmlauthorlist}

  \icmlaffiliation{yyy}{IonQ Inc., 4505 Campus Drive, College Park, MD 20740, USA}

  \icmlcorrespondingauthor{Oliver Knitter}{oliver.knitter@ionq.co}

  %\icmlkeywords{Machine Learning, ICML}

  \vskip 0.3in
]



\printAffiliationsAndNotice{}  % no special notice (required even if empty)

\begin{abstract}
  For quantum machine learning, the exact boundary between classical and quantum advantage is still poorly understood. Direct comparison between quantum neural networks (QNNs) and existing classical models, which encompass fundamentally different function classes, often fails to provide broader insight into the difference between the two. Inspired by the techniques of Neural Quantum States and Random Fourier Features, this work introduces Neural Fourier Surrogates (NFS), a stochastic classical neural network architecture for efficiently learning coefficients over the same finite Fourier series support as quantum neural networks. Testing on a selection of tabular benchmark datasets, we find that NFS is an effective classifier architecture broadly competitive with established classical baselines, including a comparable Random Fourier Features model, and possessing comparable performance to data-reuploading QNNs; combined with additional analysis comparing the learned Fourier spectra of QNNs and NFS on synthetic data, these results establish NFS as a natural classical baseline for evaluating QNN performance.
\end{abstract}

\section{Introduction}
\label{sec:intro}

Despite the considerable progress made toward the exploration of quantum machine learning as a means for producing near-term quantum advantage, an equally daunting number of unanswered questions remain. While the fundamentals of quantum neural networks (QNNs) are well understood, the exact nature of their performance, particularly the boundary delineating where they can outperform classical models, is not. Training on noisy intermediate-scale quantum (NISQ) devices is inherently difficult due to the high noise levels \cite{cerezo2021variational}, and these effects amplify issues caused by barren plateaus \cite{mcclean2018barren}. As a result, most QNNs are assessed using small datasets, executed on classical simulators, against simple classical baselines, a framework with documented systemic issues \cite{bowles2024better}; additional concerns exist as to whether the baselines used are substandard representations of classical performance
\cite{mingard2024exploiting}.

Basing QNN performance on direct comparisons with these classical baselines is particularly difficult because many of these baseline models represent wholly different function classes than QNNs, making it difficult to assess whether a QNN's performance against a specific classical model is indicative of its performance against all classical models. Conversely, even in situations where classical baselines outperform QNNs, this architectural heterogeneity still makes these results difficult to interpret. Theoretical results exist demonstrating general circumstances under which quantum models can outperform classical ones \cite{thabet2026quantum}, but as these conditions are necessary but not sufficient, they serve to further highlight the need for well-informed tangible points of comparison.

To bridge this gap, we propose classical baselines that operate on a more similar conceptual level to QNNs. Data-reuploading QNNs \cite{perez2020data} specifically model finite Fourier series whose frequency domains are determined by the encoding Hamiltonians and the number of reuploads \cite{schuld2021effect}, which is conceptually distinct from most classical baselines currently in use. Furthermore, QNN structure imposes a strong inductive bias on the finite Fourier series space \cite{mhiri2025constrained}, which the classical baselines also do not share. Taking these factors into consideration, we introduce Neural Fourier Surrogates (NFS), a novel classical neural network architecture that models the same finite Fourier series space as a data-reuploading QNN with angle encoding. These two function spaces are identical when considering QNNs with arbitrary measurement observables, but in practice, data-reuploading QNNs typically utilize a fixed measurement basis, which constrains them to a strict subset of achievable Fourier coefficients \cite{schuld2021effect}. NFS, in direct contrast, does not inherit this restriction.

Taking inspiration from Neural Quantum States (NQS), which uses classical neural nets as trial wavefunction ans\"atze amenable to Monte Carlo sampling \cite{carleo2017solving}, and Random Fourier Features (RFF), which approximates kernels using randomly sampled Fourier frequencies \cite{rahimi2007random} and has been considered more recently as a potential dequantization pipeline \cite{landman2022classically, sahebi2025dequantization}, the design of NFS comprises several independent neural network components working in tandem. An autoregressive network learns a distribution over the space of relevant Fourier frequencies, in such a manner as to allow for efficient, exact Monte Carlo sampling. The value associated with each sample is determined by a separate phase network. When multiplied by a global scaling parameter, the sample mean of these outputs approximates the desired finite Fourier series.

Under the autoregressive sampling scheme, Fourier series values may be computed without explicitly constructing every coefficient across the frequency support, stochastically overcoming the same curse of dimensionality bypassed by the QNN. Gradients of NFS outputs are similarly approximable, in a manner amenable to automatic differentiation. Through our suite of small-scale experimental benchmarks, designed to assess the fundamental performance of NFS with respect to both classical baselines and QNNs, we find that NFS is competitive with standard classical baselines on small-to-medium scale tabular data, qualitatively matches QNN performance on similar data, and serves as a way to gain insight into the Fourier spectra learned by QNNs, as demonstrated using synthetic data. Altogether, these results indicate NFS is a strong candidate as a baseline for evaluating QNN performance in future lines of research. The primary contributions of this work are as follows:
\begin{enumerate}
    \item Neural Fourier Surrogates, a novel stochastic neural net model for learning finite Fourier series.
    \item Demonstration that NFS can compete with 6 classical baselines on 15 benchmark tabular datasets
    \item Direct comparison of scaling performance between NFS and a data-reuploading QNN with respect to training set size and model width
    \item A synthetic data analysis of the ability for NFS to learn the Fourier spectrum of a QNN
\end{enumerate}

\begin{figure}[tbp]
    \centering
\includegraphics[width=0.95\columnwidth]{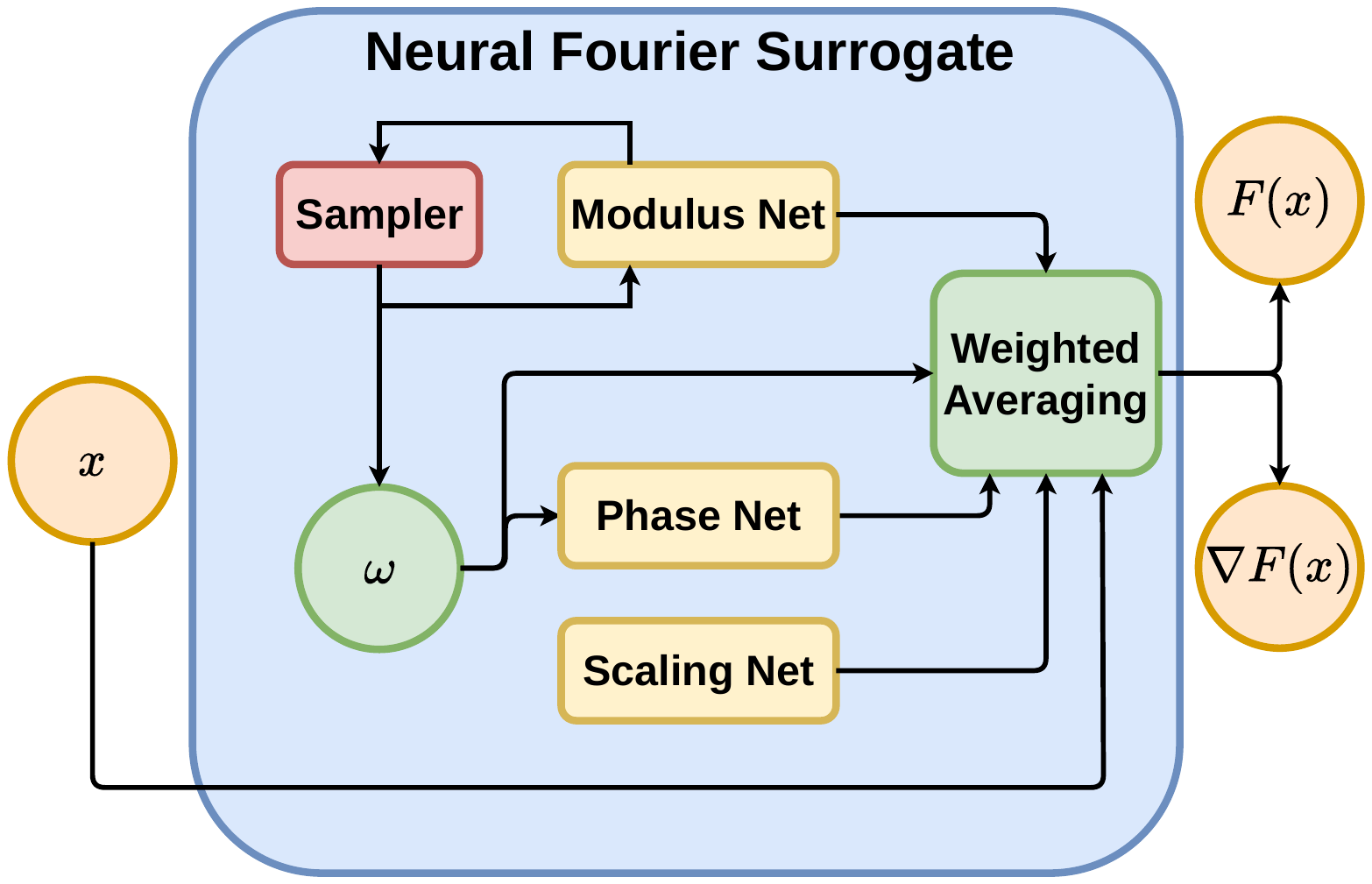}
    \caption{Basic architecture of an NFS. A sampler iteratively queries the modulus to generate a weighted set of frequencies. The modulus, phase, and scaling values are combined with the frequencies and model inputs to produce weighted averages approximating both the model output values and the corresponding gradients.}
    \label{fig:nfs_diagram}
\end{figure}

\section{Related Work}
\label{sec:related_work}

The primary influence for this work is Neural Quantum States (NQS), which uses classical neural networks to model qubit states implicitly \cite{carleo2017solving}. More specifically, this work takes inspiration from fundamental concepts of autoregressive NQS \cite{sharir2020deep}, a framework where transformers \cite{bennewitz2022neural, wu2023nnqs} or similar recurrent models \cite{hibat2020recurrent, knitter2025retentive} model qubit ans\"atze that permit efficient Monte Carlo sampling. The primary distinction between NQS and this architecture is that NQS explicitly models a single quantum state and its Born distribution, while NFS models the behavior of an entire finite Fourier series.

The \textit{classical surrogate} approach of \cite{schreiber2023classical} models a trained QNN by explicitly learning its Fourier coefficients, but this method requires enumerating across the exponentially large frequency space. At the same time, Random Fourier Features (RFF) \cite{rahimi2007random} has recently been explored as a potential avenue for dequantization \cite{landman2022classically, sahebi2025dequantization, havlivcek2019supervised}, showing that random sampling of Fourier frequencies can be used to classically mimic QNN functionality. This work is based on the fact that data-reuploading QNNs \cite{perez2020data}, where classical features are repeatedly encoded into the QNN circuit via an encoding Hamiltonian, can be modeled mathematically as a finite Fourier series \cite{schuld2021effect}. RFF samples frequencies from a fixed, model-defined distribution; in contrast, NFS learns a distribution over Fourier frequencies, allowing for greater model expressiveness.

Concurrent lines of theoretical research have explored the expressivity constraints of QNNs \cite{mhiri2025constrained} and the boundary between regions where these models and their classical surrogates outperform each other \cite{thabet2026quantum}. Our work establishes a tangible example of a classical surrogate model likely operating near this boundary.

\begin{table*}[t]
\centering
\caption{Test accuracy (mean $\pm$ std, 5 seeds) at full training size. \textbf{Bold} indicates the best result per row. The dataset name, ``B. Cancer,'' refers to Breast Cancer, and the name, ``CTG,'' refers to Cardiotocography.}
\label{tab:rq1}
\setlength{\tabcolsep}{2pt}
\footnotesize
\begin{tabular}{l r c c c c c c c}
\toprule
Dataset & $n_{\mathrm{train}}$ & NFS & MLP & RFF & SVM & RF & XGBoost & LightGBM \\
\midrule
Iris & 97 & 0.940$\pm$0.03 & 0.947$\pm$0.03 & 0.907$\pm$0.07 & \textbf{0.953}$\pm$0.03 & 0.933$\pm$0.03 & 0.940$\pm$0.04 & 0.927$\pm$0.03 \\
Wine & 115 & 0.983$\pm$0.02 & 0.972$\pm$0.02 & 0.978$\pm$0.01 & \textbf{0.989}$\pm$0.02 & 0.978$\pm$0.02 & 0.972$\pm$0.03 & 0.967$\pm$0.04 \\
B. Cancer & 369 & 0.954$\pm$0.01 & 0.953$\pm$0.02 & 0.947$\pm$0.02 & 0.963$\pm$0.02 & 0.956$\pm$0.02 & 0.958$\pm$0.02 & \textbf{0.970}$\pm$0.01 \\
Digits & 1,167 & 0.943$\pm$0.01 & 0.969$\pm$0.01 & 0.959$\pm$0.01 & \textbf{0.978}$\pm{<}0.01$ & 0.970$\pm$0.01 & 0.952$\pm$0.01 & 0.967$\pm$0.01 \\
Dry Beans & 8,846 & 0.917$\pm$0.01 & \textbf{0.928}$\pm{<}0.01$ & 0.923$\pm{<}0.01$ & 0.925$\pm{<}0.01$ & 0.919$\pm{<}0.01$ & 0.924$\pm{<}0.01$ & 0.923$\pm$0.01 \\
Abalone & 2,714 & 0.451$\pm$0.01 & \textbf{0.454}$\pm$0.01 & 0.450$\pm$0.01 & 0.443$\pm$0.01 & 0.435$\pm$0.01 & 0.441$\pm$0.02 & 0.428$\pm$0.01 \\
Ecoli & 217 & 0.847$\pm$0.04 & \textbf{0.871}$\pm$0.04 & 0.865$\pm$0.06 & \textbf{0.871}$\pm$0.05 & 0.865$\pm$0.05 & 0.824$\pm$0.04 & 0.838$\pm$0.05 \\
Glass & 138 & 0.567$\pm$0.08 & 0.712$\pm$0.05 & 0.633$\pm$0.06 & 0.688$\pm$0.10 & \textbf{0.791}$\pm$0.05 & 0.721$\pm$0.07 & 0.749$\pm$0.04 \\
Vehicle & 549 & 0.744$\pm$0.03 & \textbf{0.828}$\pm$0.02 & 0.747$\pm$0.01 & 0.765$\pm$0.02 & 0.734$\pm$0.03 & 0.739$\pm$0.03 & 0.736$\pm$0.02 \\
Segment & 1,501 & 0.927$\pm$0.02 & 0.968$\pm$0.01 & 0.921$\pm$0.01 & 0.938$\pm$0.01 & 0.973$\pm$0.01 & 0.979$\pm$0.01 & \textbf{0.983}$\pm$0.01 \\
Yeast & 964 & 0.589$\pm$0.04 & 0.593$\pm$0.02 & 0.615$\pm$0.04 & 0.621$\pm$0.03 & \textbf{0.632}$\pm$0.04 & 0.601$\pm$0.03 & 0.591$\pm$0.02 \\
Satimage & 4,179 & 0.887$\pm$0.01 & 0.899$\pm$0.01 & 0.897$\pm$0.01 & 0.894$\pm$0.01 & 0.914$\pm$0.01 & 0.913$\pm$0.01 & \textbf{0.919}$\pm$0.01 \\
Pendigits & 7,144 & 0.964$\pm$0.03 & 0.992$\pm{<}0.01$ & 0.986$\pm{<}0.01$ & \textbf{0.993}$\pm{<}0.01$ & 0.989$\pm{<}0.01$ & 0.986$\pm{<}0.01$ & 0.991$\pm{<}0.01$ \\
Letter & 13,000 & 0.822$\pm$0.03 & 0.927$\pm{<}0.01$ & 0.889$\pm$0.01 & 0.938$\pm{<}0.01$ & 0.958$\pm{<}0.01$ & 0.941$\pm{<}0.01$ & \textbf{0.961}$\pm{<}0.01$ \\
CTG & 1,381 & 0.986$\pm{<}0.01$ & \textbf{0.989}$\pm{<}0.01$ & 0.983$\pm{<}0.01$ & 0.987$\pm{<}0.01$ & 0.986$\pm$0.01 & 0.988$\pm{<}0.01$ & 0.988$\pm{<}0.01$ \\
\bottomrule
\end{tabular}
\end{table*}

\section{Methodology}

\subsection{Neural Representation of Fourier Series}
\label{sec:stochastic_representation}

Any QNN that performs Hamiltonian encoding of classical data $x\in X\subset \mathbb{R}^{d}$, specifically unitary encoding of the form $U(x) = \exp\paren{ixH}$ for some Hamiltonian $H$ into a parameterized quantum circuit $\ket{\psi_\theta(x)}$, and outputs an expectation $F_\theta(x)=\ip{\psi_\theta(x)|O|\psi_\theta(x)}$ with respect to an observable $O$, is equivalent to a Fourier series \begin{equation}F_\theta(x) = \sum_{\omega\in\Omega} f_\theta(\omega)e^{i\omega\cdot x},\label{eqn:fourier_decomp}\end{equation} where the frequency support $\Omega\subset \mathbb{Z}^d$ is determined by the dimension $d$ of $X$ and the number $L$ of encoding re-uploads \cite{schuld2021effect}.  We suggest using a neural network that maps frequencies to Fourier coefficients \[\omega\mapsto f_\theta(\omega),\] mirroring how NQS ans\"atze map qubit spin configurations to statevector entries. To make the computation of $F_\theta$ tractable, we stochastically approximate it via random frequency sampling; splitting $f_\theta$ into its modulus $\abs{f_\theta}:\Omega\rightarrow [0,\infty)$ and phase $\phi_\theta:\Omega\rightarrow\bracket{-\pi,\pi}$ gives \begin{align*}F_\theta(x) &= \sum_{\omega\in\Omega} \abs{f_\theta(\omega)}e^{i\paren{\phi_\theta(\omega) + \omega\cdot x}}\\
&= \norm{f_\theta}_1\E_{\omega\sim \abs{f_\theta}}\bracket{e^{i\paren{\phi_\theta(\omega) + \omega\cdot x}}}\numberthis
    \label{eqn:random_variable}
\end{align*}

where $\norm{f_\theta}_1=\sum_{\omega\in\Omega} \abs{f_\theta(\omega)}$ and the expectation $\omega\sim\abs{f_\theta}$ is taken over the 1-normalized modulus. In this paper, expectations over $\omega$ are taken with respect to this distribution unless otherwise noted. Note the imaginary part in (\ref{eqn:random_variable}) has zero expected value. Representing $\abs{f_\theta}$ and $\phi_\theta$ with neural networks, we can stochastically approximate $F_\theta$ by collecting samples from (\ref{eqn:random_variable}) and averaging them. To facilitate training, we apply the log-derivative trick to (\ref{eqn:fourier_decomp}) to find a stochastic estimator for $\nabla_\theta F_\theta$:

\begin{align*}
    \nabla_\theta F_\theta(x) &= \sum_{\omega\in\Omega} e^{i\omega\cdot x}\nabla_\theta e^{\log f_\theta(\omega)}\\
    &= \sum_{\omega\in\Omega} e^{i\omega\cdot x} e^{\log f_\theta(\omega)}\nabla_\theta \log f_\theta(\omega)\\
    &= \sum_{\omega\in\Omega} \abs{f_\theta(\omega)}e^{i\paren{\phi_\theta(\omega) + \omega\cdot x}} \nabla_\theta\log f_\theta(\omega)\\
    &= \norm{f_\theta}_1\E_{\omega}\bracket{e^{i\paren{\phi_\theta(\omega) + \omega\cdot x}}\nabla_\theta\log f_\theta(\omega)}\numberthis\label{eqn:log_derivative}
\end{align*} where again only the real part of this expectation is needed. A similar calculation yields the gradient of $F_\theta$ with respect to the model inputs:

\begin{equation}
    \nabla_{x}F_\theta(x) = \norm{f_\theta}_1 \mathbb{E}_{\omega}\bracket{e^{i(\phi_\theta(\omega) + \frac{\pi}{2}+\omega\cdot x)}\omega}.
    \label{eqn:input_derivative}
\end{equation} Sample means approximating (\ref{eqn:log_derivative}) and (\ref{eqn:input_derivative}) allow for NFS to operate within any automatic differentiation pipeline that allows for custom gradient calculation. To reduce sampling cost, we use the same frequency sample batch to estimate both $F_\theta(x)$ and its gradients, which we note introduces a covariance-derived bias inversely proportional to the sample batch size. As discussed in section \ref{sec:experiments}, the autoregressive sampling structure of NFS permits sample batch sizes for which this bias becomes negligible.

\begin{figure*}[t]
    \centering
    \includegraphics[width=0.85\textwidth]{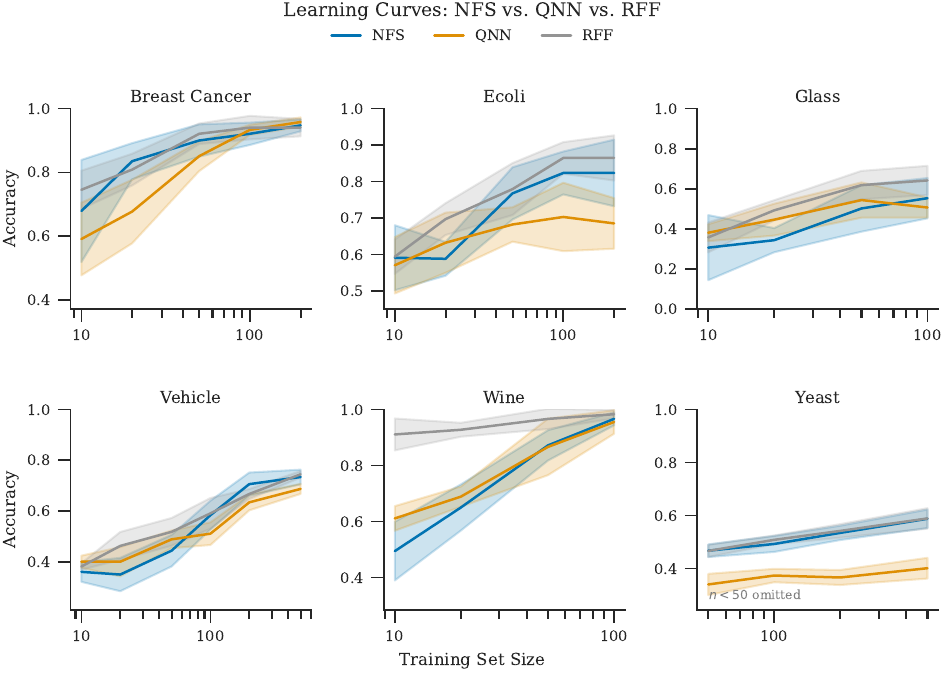}
    \caption{Test accuracy, as a function of training set size, for a Neural Fourier Surrogate with encoding depth $L=4$, an equivalent QNN, and a Random Fourier Features (RFF) baseline on six benchmark datasets. Shaded regions show $\pm 1$ standard deviation across 5 seeds.}
    \label{fig:nfs_qnn_learning_curves}
\end{figure*}

\subsection{Neural Network Architecture}
\label{sec:implementation}

The general NFS architecture determined by this paper is depicted in Figure \ref{fig:nfs_diagram}, though we do note other alternative implementations are likely possible. For effective use as a QNN surrogate, the neural networks modeling $\abs{f_\theta}$ and $\phi_\theta$ must permit the following key features:

\begin{enumerate}
    \item An efficient sampling scheme for $\omega\sim\abs{f_\theta}$.

    \item A guarantee of conjugate symmetric Fourier coefficients, equivalent to saying that $\abs{f_\theta}$ is even and $\phi_\theta$ is odd with respect to $\omega$, so as to ensure (\ref{eqn:random_variable}) is always real-valued.

    \item Generalization to multi-dimensional model outputs.
\end{enumerate}

To address feature 1, we implement $\abs{f_\theta}$ as a transformer-based model \cite{vaswani2017attention} similar to the autoregressive NQS ans\"atze first developed for \textit{ab initio} quantum chemistry \cite{barrett2022autoregressive}, specifically the NNQS-Transformer \cite{wu2023nnqs}. Following the design principle of these NQS models, we use a shallow feedforward network for $\phi_\theta$. Given a frequency $\omega = \paren{\omega_1,\dots,\omega_d}$, $\abs{f_\theta}$ returns a sequence of nested conditional probabilities \begin{equation}
    \abs{f_\theta(\omega)} = \paren{\operatorname{Pr}(\omega_1), \operatorname{Pr}(\omega_2|\omega_1),\dots,\operatorname{Pr}(\omega_d|\omega_{j<d})}.
\end{equation} Each of these conditional probabilities has a relatively small support, in this case $[-L,\dots,L]$, and their product $\prod_i \operatorname{Pr}\paren{\omega_i|\omega_{j<i}}$ gives a normalized joint probability for $\omega$. With these outputs, autoregressive sampling from $\Omega$ can be done in an efficient, easily parallelized procedure \cite{zhao2021overcoming}. This model does depend on a user-specified ordering of the frequency variables within $\omega$. Though the correlations between input features may favor certain orderings over others, we default to a random ordering and trust the model to learn from it.

Since the moduli of the coefficients in (\ref{eqn:fourier_decomp}) are not normalized, we use a third network that takes no input and generates the 1-norm $\norm{f_\theta}_1$ of these coefficients, allowing the NFS full expressivity within the finite Fourier series domain while retaining the benefits of exact sampling from normalized probabilities. We note that this scaling network produces the same output regardless of frequency and in practice can comprise a single positive trainable parameter per output dimension.

For feature 2 to hold, it must be that $\abs{f_\theta(-\omega)} = \abs{f_\theta(\omega)}$ for all $\omega$. Through appropriate masking of the conditional probabilities outputted by $\abs{f_\theta}$, we restrict sampling to the subspace of $\Omega$ whose first entries are all nonzero. We can incorporate the contribution of negative frequencies into the formulas from section \ref{sec:stochastic_representation} by doubling the contribution of nonzero frequencies. This masking has the added benefit of alleviating some of the compute cost associated with sampling. To reduce this cost further, we utilize importance sampling gradient optimization (ISGO), which reuses frequency samples across multiple gradient updates in an unbiased way and has been proven effective for NQS \cite{yang2020deep}.

It remains to ensure that $\phi_\theta(\omega)$ is an odd function, which we achieve by anti-symmetrizing the output of the neural network $g$ we use to encode $\phi_\theta$: \begin{equation}
    \phi_\theta(\omega) = \frac{g_\theta(\omega) - g_\theta(-\omega)} 2.
\end{equation} We augment the ability for $\phi_\theta$ to learn meaningful information from integer frequency inputs by utilizing a simple NeRF-style \cite{mildenhall2020nerf} positional encoding \begin{equation}
    \bracket{
        \cos\paren{\omega}, \sin\paren{\omega}, \dots, \cos\paren{2^{k-1}\omega}, \sin\paren{2^{k-1}\omega}},
\end{equation} where $k$ is some user-defined multiplier that sets the expressiveness of the embedding.

Turning our attention to feature 3, the NFS framework naturally generalizes to handle multidimensional outputs, representing measurements of the same quantum circuit by multiple observables, by extending (\ref{eqn:fourier_decomp}) to a vector form: \begin{align*}
    F_\theta(x) &= \paren{F_\theta^1(x),\dots,F_\theta^n(x)}\\
    &= \sum_{\omega\in\Omega}\paren{f_\theta^1(\omega),\dots,f_\theta^n(\omega)}e^{i\omega\cdot x}
    \numberthis
\end{align*}

Each coefficient $f_\theta^j$ represents a separate Fourier series with the same support. From here, the decomposition of each $f_\theta^j$ into modulus and phase components naturally induces $n$ unnormalized probability distributions over $\Omega$, which may be sampled in parallel. In practice, we use the same $\abs{f_\theta}$ and $\phi_\theta$ networks to encode all Fourier series simultaneously, with a projection head in the final layer determining the output for each dimension. This design is motivated by the fact that all Fourier series corresponding to a multi-output QNN typically share the same underlying circuit. For this study, we use NFS outputs as multiclass classification logits trained in a supervised framework using a mean squared error loss.

\begin{figure*}[t]
    \centering
    \includegraphics[width=0.85\textwidth]{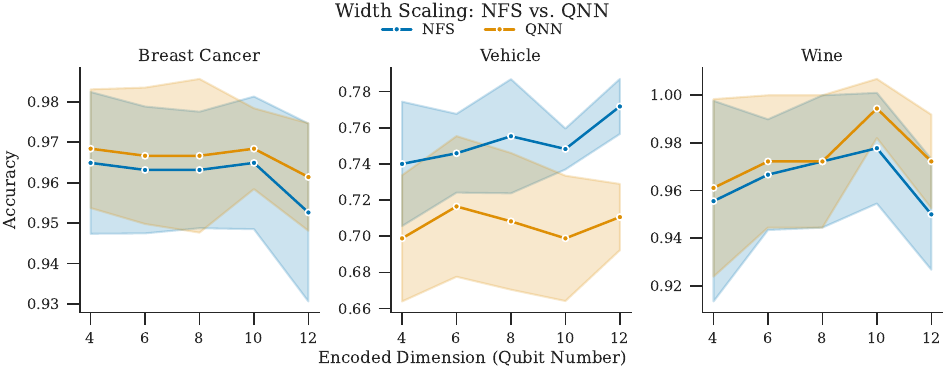}
    \caption{Test accuracy as a function of encoded dimension (qubit number) for an NFS and a QNN with encoding depth $L=1$, trained on up to 500 samples across 3 datasets. Shaded regions show $\pm 1$ standard deviation across 5 seeds.}
    \label{fig:nfs_qnn_width_scaling}
\end{figure*}

\subsection{Gaussian Prior Regularization}
\label{sec:gaussian_prior}

It is known that data-reuploading QNNs place an implicit prior on the magnitudes of their Fourier coefficients. The variance of $f_\theta(\omega)$ over random weight initializations decreases with $\abs{\omega}$, as there exist fewer paths to construct higher frequencies from differences of encoding eigenvalues \cite{schuld2021effect}. NFS does not inherently have this inductive bias, and we explicitly incorporate it by regularizing the learned frequency distribution toward a Gaussian $G(\omega)\propto\exp \paren{\frac{-\norm{\omega}^2}{2L}}$, as a proxy for this path count structure: \begin{equation}
    \operatorname{KL}\paren{\abs{f_\theta}\parallel G} = -H_{\omega} + \mathbb{E}_{\omega} \bracket{\frac{\norm{\omega}^2}{2L}} + \log Z
\end{equation} where $H_\omega$ represents the entropy of $\omega\sim\abs{f_\theta}$ and $Z$ is the normalization constant of $G$. This regularization is similar to variational neural annealing (VNA) \cite{hibat2021variational}, and is implemented analogously to how VNA is applied for NQS \cite{knitter2025retentive}.

\begin{figure*}[t]
    \centering
    \includegraphics[width=0.85\textwidth]{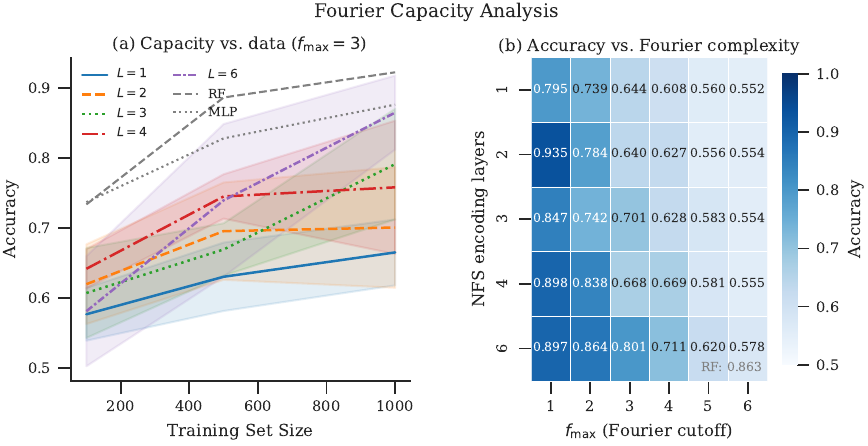}
    \caption{(a) NFS test accuracy as a function of training set size on a synthetic Fourier target function with cutoff $f_{\mathrm{max}} = 3$. Curves shown for encoding depth choices between 1 and 6, alongside Random Forest and MLP baselines. (b) NFS test accuracy across all combinations of encoding depth and Fourier cutoff $L,f_{\textrm{max}}\in\braces{1,\dots,6}$, averaged across 5 seeds. Higher choices of $L$ provide diminishing returns beyond the target's spectral complexity.}
    \label{fig:nfs_fourier_capacity}
\end{figure*}

\section{Experiments}
\label{sec:experiments}

In this section, we assess the basic capabilities of NFS by training it on a set of 15 benchmark datasets sourced from the OpenML \cite{vanschoren2014openml} and UC Irvine \cite{asuncion2007uci} repositories. These datasets cover multiple instances of multiclass tabular classification for a variety of topics and sample sizes. Using a 65/15/20 train/validation/test split, we generate problem instances spanning 97 to 13,000 training samples. For each dataset, all features were normalized using training data statistics. Unless otherwise noted, all test results are reported as means with standard deviations across 5 random seeds, alongside baselines taken from a variety of comparable architectures depending on the exact analysis being performed.

The NFS used in these tests consists of a transformer-based modulus network and a feedforward phase network. We compare the performance of this model with that of an angle-encoded data-reuploading QNN  \cite{perez2020data} implemented in PennyLane \cite{bergholm2018pennylane}. This QNN uses a ladder structure of $Y$-axis single qubit rotations and $\operatorname{CNOT}$ gates, followed by $Z$-basis qubit measurements, all simulated noiselessly. Both architectures are combined with a linear encoder layer for improved feature extraction, and are trained using AdamW \cite{loshchilov2017decoupled} for 200 epochs, using a batch size of 32 and a cosine annealing scheduler beginning at $10^{-3}$. NFS training also incorporates a fixed Gaussian prior regularization as discussed in section \ref{sec:gaussian_prior}. The learning rate and regularization coefficient, alongside other key hyperparameters, were determined according to a tuning analysis we discuss in section \ref{sec:ablation}, while the number of NFS non-unique frequency samples per gradient step is scheduled geometrically from one thousand to one trillion, which progressively reduces gradient estimation bias and variance as training proceeds. For classical baselines, we consider multilayer perceptron (MLP), support vector machine (SVM) \cite{cortes1995support}, and random forest (RF) \cite{breiman2001random} models trained using the default scikit-learn hyperparameters \cite{pedregosa2011scikit}, alongside XGBoost
\cite{chen2016xgboost} and LightGBM
\cite{ke2017lightgbm} trained with their respective package defaults. We additionally include an RFF baseline implemented through the scikit-learn RBFSampler, with 500 random features and the bandwidth term $\gamma$ chosen from $\braces{0.001, 0.01, 0.05, 0.1}$ using the validation set, as a direct point of comparison between NFS and a fixed-distribution Fourier feature model.

\subsection{Full Comparison on Tabular Benchmark Data}
\label{sec:rq1}

Before assessing its capacity as a quantum surrogate, we first examine the baseline performance of NFS as a supervised learner by testing it across all 15 benchmark datasets in comparison with all classical baselines. In these instances, we do not necessarily expect NFS to exceed baseline performance since these classical models are particularly well-suited for tabular data \cite{grinsztajn2022tree}. Within our results presented in Table \ref{tab:rq1}, we find that despite no evidence of superior performance, NFS is broadly competitive across all benchmarks. NFS performance does generally degrade on the higher-dimensional benchmark datasets; all tests were performed with a qubit dimension $d=10$ to match the width of the NISQ-scale QNN from section \ref{sec:qnn_comparison}, which could offer an explanation for this behavior. A limited study of the effect of $d$ on NFS and QNN performance is included in section \ref{sec:qnn_comparison}.

\begin{table}[t]
\centering
\caption{A small hyperparameter sensitivity analysis for NFS. Accuracy is determined by the mean accuracy, for training set sizes $n\in\braces{100,500}$, across all datasets in Table \ref{tab:rq1} with at least 500 training samples, and 5 seeds. $\dagger$ marks the default value used in all other experiments.}
\label{tab:ablation}
\small
\begin{tabular}{ll cc}
\toprule
Parameter & Value & $n{=}100$ & $n{=}500$\\
\midrule
\texttt{encoding\_depth} & 1 & 0.724 & 0.806\\
 & 2 & 0.700 & 0.794\\
 & 4$^\dagger$ & 0.684 & 0.794\\
 & 8 & 0.608 & 0.763\\
\addlinespace[2pt]
\texttt{embed\_dim} & 8 & 0.696 & 0.765\\
 & 16$^\dagger$ & 0.711 & 0.783\\
 & 32 & 0.726 & 0.768\\
 & 64 & 0.723 & 0.780\\
\addlinespace[2pt]
\texttt{max\_uniqs} & 100 & 0.658 & 0.740\\
 & 500 & 0.712 & 0.770\\
 & 2k & 0.697 & 0.771\\
 & 10k$^\dagger$ & 0.705 & 0.757\\
\addlinespace[2pt]
\texttt{Gauss\_coeff} & 0 & 0.634 & 0.737\\
 & 0.25$^\dagger$ & 0.706 & 0.775\\
 & 0.75 & 0.705 & 0.767\\
 & 2 & 0.703 & 0.749\\
\addlinespace[2pt]
\texttt{learning\_rate} & 1e-04 & 0.342 & 0.526\\
 & 5e-04 & 0.656 & 0.762\\
 & 1e-03$^\dagger$ & 0.719 & 0.772\\
 & 5e-03 & 0.751 & 0.791\\
\bottomrule
\end{tabular}
\end{table}

\subsection{Learning Efficiency Relative to QNNs}
\label{sec:qnn_comparison}

Since NFS is explicitly built on the same type of finite Fourier frequency structure as QNNs, we now move to a direct comparison in performance between the two on a selection of benchmark datasets from section \ref{sec:rq1}: Breast Cancer, \textit{E. coli}, Glass, Vehicle, Wine, and Yeast. For each of these datasets we sampled selections of training data points between 10 and $\sim1000$ in order to capture the performance of these models at a variety of training set sizes. In Figure \ref{fig:nfs_qnn_learning_curves} we present as a function of training set size the final test accuracy, averaged over 5 seeds, achieved by the NFS and QNN models, both with an encoding depth of $4$. These results indicate the two models exhibit consistently similar performance relative to each other and to the RFF baseline, which for most datasets displays a qualitatively similar level of performance to the NFS and QNN, as also seen in section \ref{sec:rq1}.

While NFS generally outperforms the QNN in these experiments, this should not be taken as a uniform sign of advantage. As discussed in section \ref{sec:gaussian_prior}, the choice of a fixed measurement observable does introduce a greater implicit bias and constrain the expressivity of the QNN beyond the general finite Fourier series structure. The addition of more complicated, or tunable, observables could close the gap. 

Moreover, the results in Figure \ref{fig:nfs_qnn_width_scaling} generally show QNN outperforming NFS. In contrast with Figure \ref{fig:nfs_qnn_learning_curves}, where all tests were performed using a qubit dimension of 10, here we explore the effect of varying qubit dimension on NFS and QNN performance on 3 of the benchmark datasets, when holding the training set size fixed at $\sim500$. The qubit dimension is varied from 4 to 12; since all three datasets have at least 12 features, we can ensure the encoder layer never maps into a larger space. Both models are fairly insensitive to the change in qubit dimension, indicating the primary performance bottleneck lies in the dimension reduction. The fundamental architectural differences between the two models do not allow for a fair parameter-matched performance comparison between them, and this capacity sweep serves as the closest feasible analog to such a comparison. We do note the similarity in qualitative behavior between the two models across three plots. Altogether, these results appear to support the use of NFS as a classical point of reference when evaluating QNN performance. At the same time, the fact that NFS operates within a slightly more expressive function class than the QNN used here does warrant a closer examination into the ability of NFS to learn QNN-specific Fourier series.

\begin{figure}[t]
    \centering
    \includegraphics[width=0.85\columnwidth]{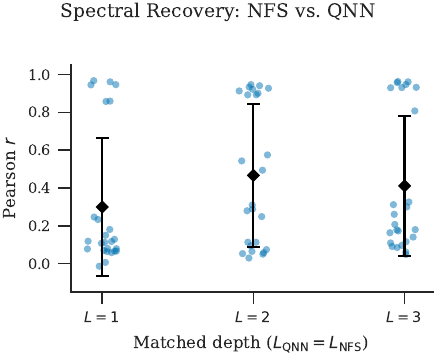}
    \caption{Pearson $r$ between the learned frequency distribution of an NFS and the Fourier spectrum of a fixed random QNN, at matched encoding depths, averaged over 25 trials. Error bars show the mean $\pm 1$ standard deviation. NFS was trained to match randomly sampled QNN output. The bimodal distribution indicates NFS either recovers the QNN distribution or learns an altogether different one.}
    \label{fig:nfs_coeff_recovery}
\end{figure}

\subsection{Fourier Capacity Analysis of NFS}

Understanding that NFS performs similarly to QNNs on basic supervised learning tasks, we also wish to evaluate how the model capacity of an NFS, as represented by its encoding depth $L$, interacts with data complexity. To that end, we fit the NFS to synthetic data generated by querying a randomly initialized QNN with a known encoding depth $f_{\textrm{max}}$, allowing us to control both the size and complexity, as represented by the Fourier frequency spectrum of the dataset. We present our findings in Figure \ref{fig:nfs_fourier_capacity}. In part (a), we consider datasets of size 100, 500, and 1000 with Fourier complexity $f_{\textrm{max}}=3$, using them to train NFS models with encoding depth $L$ chosen between 1 and 6. We observe that as $L$ increases, the ability of the NFS to replicate the QNN output generally grows as well, though the $L=6$ case appears to indicate that models with higher encoding depth require more training data to be effective on data of lower complexity, likely due to overfitting.

In part (b), we examine how test accuracy is affected as we vary both $L$ and $f_{\textrm{max}}$ under a fixed training data size of 500. Our results depict the two fundamental trends one would expect: increasing dataset complexity reduces model performance, and increasing encoding depth improves it. These relationships corroborate the claim that encoding depth serves as a primary determining factor of model capacity. We note that at sufficiently high Fourier complexity, the increase in model capacity produces only a marginal increase in performance; combined with the results of part (a), this behavior appears to indicate that higher capacity models require more data to train effectively.

These NFS models were trained specifically to match their outputs to those of the QNNs that generated the data, but an interesting phenomenon appears when comparing the models more closely. Matching the model encoding depth $L$ to the complexity of the generated data, we compute the Pearson $r$ value between the frequency distribution of the trained NFS and that of the data-generating QNN. In Figure \ref{fig:nfs_coeff_recovery} we show the distribution over 25 trials of these values for each of the encoding depths 1, 2, and 3. In all cases the distributions are bimodal, clustering around both 1 and 0. This phenomenon can be explained by the greater expressiveness of the NFS relative to the QNN used here, since the latter employs a fixed observable. The NFS either converges to the same spectral solution, or it finds a different suitable coefficient set that is inaccessible to the QNN.

\subsection{Hyperparameter Tuning}
\label{sec:ablation}

We performed a basic hyperparameter sensitivity analysis to determine the exact configuration used in the other experiments, evaluating the effect of the NFS encoding depth; the transformer embedding dimension; the maximum number of unique samples generated at each training step, which is the true measure of compute cost under the parallelized sampling structure NFS inherits from autoregressive NQS \cite{zhao2021overcoming}; the Gaussian prior regularization coefficient; and the learning rate. Due to the stochastic sampling, NFS training necessitates a greater computational cost than the classical baselines discussed in section \ref{sec:rq1}, which this work does not explore in detail as the NFS is primarily intended to serve as an interpretive surrogate for quantum models, and not a competitor to other classical methods.

We tested these hyperparameter configurations on the benchmark datasets from section \ref{sec:rq1} at training set sizes of 100 and 500, and we present the average test accuracy for each data point in Table \ref{tab:ablation}. Our findings stress the importance of a sufficiently high learning rate and the Gaussian prior regularization, and that lower model capacity performs better with smaller training data size. Otherwise, NFS is generally robust to reasonable choices of the other hyperparameters.

\section{Conclusion}
\label{sec:conclusion}

Neural Fourier Surrogates comprise a novel stochastic neural network architecture that learns coefficients for the same type of complex finite Fourier series modeled by data-reuploading Hamiltonian encoding quantum neural networks, spanning a coefficient space theoretically achievable for the entire QNN function class, but slightly larger than what typical QNN architectures can reach. Using a transformer-based model, we have demonstrated that NFS exhibits similar performance to a QNN across small-to-medium tabular datasets and is competitive with, though not superior to, classical baseline performance in this domain. NFS represents a broader class of function than a QNN with a fixed measurement observable, and as a result it does not possess the same expressivity constraints that have been demonstrated for QNNs \cite{mhiri2025constrained}.

These results position NFS as a strong classical baseline for assessing QNN performance, helping us to better understand the boundary between classical and quantum advantage in machine learning \cite{thabet2026quantum}. Useful directions for future work in this area include comparisons between NFS and QNNs that have been executed on quantum hardware \cite{knitter2026measuring}, and the exploration of whether efficient recurrent neural networks, which have been considered in the NQS space \cite{knitter2025retentive}, may help improve NFS scaling cost without sacrificing accuracy.

\subsection*{AI Use Statement}
Generative AI (Anthropic Claude) was used by the authors for purposes of project and experiment ideation, benchmark code generation, and general manuscript outlining and proofreading. Generative AI was not used for mathematical derivations, manuscript writing, theoretical development, or data generation. The authors verified all AI-generated output and take full responsibility for all content in this submission.

\subsection*{Acknowledgments}
The authors would like to thank Alex Buser for his influence on the experimental framework in this paper.

\bibliography{references}
\bibliographystyle{icml2026}

\end{document}